\documentclass[structabstract]{aa}  
\usepackage{graphicx}
\usepackage{txfonts}
\usepackage{natbib} 
\bibpunct{(}{)}{;}{a}{}{,} 
\begin{document}
   \title{WINGS Data Release: a database of galaxies in nearby clusters}

   \author{
   	A. Moretti 
	\inst{1,2}
          \and
         B.M. Poggianti 
         \inst{2}
         \and
         G. Fasano \inst{2}
         \and
         D. Bettoni \inst{2}
         \and
         M. D'Onofrio \inst{1}
         \and
         J. Fritz \inst{3}
         \and
         A. Cava \inst{4}
          \and 
          J. Varela \inst{5}     
         \and
          B. Vulcani \inst{6}
          \and 
         M. Gullieuszik \inst{2}
         \and
         W. J. Couch \inst{7}
         \and
         A. Omizzolo \inst{2}
          \and
          T. Valentinuzzi \inst{1}
         \and
         A. Dressler \inst{8}
         \and
          M. Moles \inst{5}
          \and
          P. Kj{\ae}rgaard \inst{9}
          \and
          R. Smareglia \inst{10}
          \and
          M. Molinaro \inst{10}
                    }

   \institute{
               University of Padova, Department of Physics and Astronomy "G. Galilei",
             Vicolo dell'Osservatorio, 2, Padova - Italy\\  
              \email{alessia.moretti@oapd.inaf.it}
         \and
             INAF-Osservatorio Astronomico di Padova,
              Vicolo dell'Osservatorio, 5, Padova, Italy
             \and 
             Sterrenkundig Observatorium Vakgroep Fysica en Sterrenkunde Universiteit Gent, Krijgslaan 281, S9 9000 Gent, Belgium
             \and 
             Observatoire de Gen{\`e}ve, Universit{\'e} de Gen{\`e}ve, 51 Ch. des Maillettes, 1290 Versoix, Switzerland
             \and 
             Centro de Estudios de F'sica del Cosmos de Aragon,
             Plaza San Juan, 1 -  44001 Teruel, Spain
              \and 
	    Kavli Institute for the Physics and Mathematics of the Universe (WPI), Todai Institutes for Advanced Study, the University of Tokyo, Kashiwa, 277-8582, Japan
             \and 
             Australian Astronomical Observatory, PO Box 915, North Ryde, NSW 1670 Australia
             \and 
             Carnegie Observatories, 813 Santa Barbara Street,
             Pasadena, CA 91101, USA
             \and 
             Niels Bohr Institute, Juliane Maries Vej 30, 2100 Copenhagen, Denmark
             \and 
             INAF-Osservatorio Astronomico di Trieste,
	   Via G.B. Tiepolo 11, Trieste,Italy}

   \date{Received November 21, 2013; accepted March 3, 2014}
 
  \abstract
{To effectively investigate galaxy formation and evolution, it is of paramount importance to exploit homogeneous data for large samples of galaxies in different environments.}
{The WINGS (WIde-field Nearby Galaxy-cluster Survey) project aim is to evaluate physical properties of galaxies in a complete sample of low redshift clusters to be used as reference sample for evolutionary studies. The WINGS survey is still ongoing and the original dataset will soon be enlarged with new observations. This paper presents the entire collection of WINGS measurements obtained so far.}
{We decided to make use of the Virtual Observatory (VO) tools to share the WINGS database (that will be regularly updated) with the community. In the database each object has one unique identification (WINGSID). Each subset of estimated properties is accessible using a simple cone search (including wide-field images).}
{We provide the scientific community the entire set of wide-field images. Furthermore, the published database contains photometry of 759,024 objects and surface brightness analysis for 42,275 and 41,463 galaxies in the V and B band, respectively. The completeness depends on the image quality, and on the cluster redshift, reaching on average $90\%$ at V$\lesssim 21.7$.
Near infrared photometric catalogs for 26 (in K) and 19 (in J) clusters are part of the database and the number of sources is 962,344 in K and 628,813 in J. Here again the completeness depends on the data quality, but it is on average higher than $90\%$ for J$\lesssim20.5$ and K$\lesssim19.4$.
The IR subsample with a Sersic fit comprises 71,687 objects. A morphological classification is available for 39,923 galaxies. We publish spectroscopic data, including 6,132 redshifts, 5,299 star formation histories and 4,381 equivalent widths. Finally, a calculation of local density is presented and implemented in the VO catalogs for 66,164 galaxies. The latter is presented here for the first time.}
{}


 \maketitle
%


\section{Introduction}\label{sec:intro}

The WINGS \citep[WIde--field Nearby Galaxy--cluster Survey,][]{Fasano2006} project was conceived to give a full description of galaxies in nearby clusters, and to provide a robust and homogeneous observational dataset to be used in the interpretation of galaxies in clusters at higher redshift. 

In this context, current knowledge of the systematic properties of galaxies in nearby clusters remains surprisingly limited, being largely based on just the Virgo, Coma, Fornax clusters and the Shapley supercluster \citep[e.g.]{Gavazzi2003,Merluzzi2010}. At higher redshifts, the LoCUSS survey has targeted galaxy clusters at $z \sim 0.2$, the STAGES project has studied in detail the A901/2 system at $z=0.165$ \citep{Smith2010,Gray2009}, and a large amount of high quality data for more distant clusters is continuously being gathered with the Hubble Space Telescope (HST) and large ground-based telescopes.

Nevertheless, the morphological reference for local clusters is still provided by the historical database of \citet{dressler80} based on photographic plates, giving the positions, the estimated magnitudes (down to V $\sim 16$) and the visual morphological classification for galaxies in $55$ clusters in the range $0.011 \leq z\leq 0.066$. This awkward situation can be easily understood since only with the new large format (wide-field) CCD mosaic cameras a significant number of low redshift clusters could be reasonably well mapped.
On the other hand, morphological classifications are presently available for a large number of galaxies from the Sloan Digital Sky Survey (SDSS), that has not been designed to be a cluster survey. In fact, morphological classifications are available only for the brightest sources ($g \leq 16$ in \citealp{Nair2010, Fukugita2007} and $r \leq 17$ in \citealp{Willett2013}) and are thus much less accurate than those provided by a dedicated survey \citep[see for example the EFIGI catalogue by][with detailed morphology but magnitude limit at $g=14$]{Baillard2011}.

The WINGS collaboration has started to fill the observational gap between very nearby clusters and high-redshift ones by observing in the optical bands (B and V) $76$ clusters of galaxies that span the largest possible range of cluster characteristics, as given by their X--rays properties.

WINGS clusters have been selected from three X-ray
flux-limited samples compiled from ROSAT All-Sky Survey data: the ROSAT Brightest Cluster Sample \citep[][BCS]{ebeling1998}, and its extension \citep[][eBCS]{ebeling2000} in the Northern hemisphere and the X-Ray-Brightest Abell-type Cluster sample \citep[][XBACs]{ebelingH.1996} in the Southern hemisphere.  The original WINGS sample comprises all clusters from
BCS, eBCS and XBACs with a high Galactic latitude ($|b|\geq 20$ deg) in the redshift range $0.04 < z < 0.07$. The redshift cut
and the Galactic latitude are thus the only selection criteria applied to the X-ray samples.
We refer the reader to the original paper by \citet{Fasano2006} for a description of the cluster sample (i.e. X--ray luminosities, temperatures, masses, Bautz--Morgan class distribution and so on).

The optical CCD imaging data obtained for this sample of local clusters of galaxies is called WINGS-OPT, and is the main foundation upon which the WINGS project is based. From this set of mosaic images it has been possible to construct a photometric catalog of sources suitable for spectroscopic follow--up (from now on called WINGS-SPE, see \citealp{Cava2009} for a more detailed description). Unfortunately our spectroscopic programme suffered from bad weather conditions, so the final WINGS-SPE sample contains $48$ (of the $76$) clusters, $22$ of which are in the southern sky and $26$ in the north. 

For a subsample of clusters, photometric data in the $J$ and $K$ near-infrared wavebands were added to the main photometric sample, via a dedicated programme undertaken with UKIRT. WFCAM observations were taken of 28 clusters \citep{Valentinuzzi2009}, making WINGS-NIR the largest near-infrared survey of nearby galaxy clusters in terms of areal coverage. In fact, in this redshift range, only individual clusters or small cluster samples have been studied in the literature up to now (e.g. \citealp{Pahre1999}; \citealp{Gavazzi1990}; \citealp{DePropris2003}).

Dedicated observations with the INT, BOK and LBT telescopes have also provided U band photometry for a subsample of 17 clusters \citep[see][]{Omizzolo2013}.

Finally, the project is being expanded by very wide--field observations (4 times the original WINGS area) taken with Omegacam/VST in the usual $B$ and $V$ Johnson bands  (Gullieuszik et al., in preparation) and in the Sloan u band.
A programme of follow-up spectroscopy using AAOmega/AAT is currently in progress, and we plan to release a second version of the database covering this larger area. The OMEGACAM/AAOmega dataset will extend
out to between 2 and 5 times the projected virial radius of each cluster.\footnote{In all WINGS papers the virial radius
of each cluster is taken to be equal to $R_{200}$, the radius delimiting a sphere with interior mean density
200 times the critical density, and is derived from the cluster velocity dispersion as in Cava et al. 2009).}

\section{Overview of the database}

In order to fully exploit the capabilities of the entire dataset, we decided to develop a local database of measurements, whose structure and size grows as the observations improve.
The goal of the present paper is to describe the whole WINGS dataset, as available for download and use through Virtual Observatory (VO) tools. 
We decided to use this method to make the data public, since it is becoming more and more popular, and it is relatively easy to maintain a VO mirror of our local MySQL database.

WINGS data have been published so far together with their describing papers through the  {\it Strasbourg astronomical Data Center} (Centre de Donnees astronomiques de Strasbourg, CDS) (\citealt{Varela2009}; \citealt{Cava2009}; \citealt{Valentinuzzi2009}; \citealt{Fritz2010}; Bindoni et al., in preparation). However, these catalogs can not be regularly updated. We therefore decided to make our data available using the VO, which allows a more flexible treatment (and usage) of the data itself. The registration of WINGS services to the VO repository is maintained by the IA2 team in Trieste \citep{Molinaro2012}.

The structure of the database is very simple, as it has one unique primary key, which is the WINGS ID. Each new measurement (photometry, spectra, profiles and so on) is then uploaded into the database with its own unique ID. 
The WINGS ID is a string field of 25 characters which contains the coordinates of the object (right ascension and declination in hhmmss and deg).
In order to avoid identifications that are not meaningful, each time a new measurement is made an appropriate series of routine checks are made  for possible cross-matching objects inside a given coordinates box, whose size depends on the positioning error (due, for example, to the astrometry).
We provide in Tab. \ref{tab:WINGSdb}  a summary of the different datasets presently constituting the WINGS archive, that will be described in this paper. The number in the fourth column gives the total number of objects for each dataset.

\begin{table*}
\caption{WINGS database in a nutshell: for each dataset we list in this table the section of the present paper where the dataset is described, the content of the dataset, the number of objects constituting the dataset and the original paper with the detailed analysis.}             
\scriptsize
\label{tab:WINGSdb}      
\centering               
\begin{tabular}{l l l l l}
\hline\hline              
Table name & Ref & Content & N(obj) & References\\    
\hline                        
   OPT                   &Sec. \ref{sec:opt}   & Optical V (B) photometry & 759024 & (1) \\      
   GASPHOTV     & Sec. \ref{sec:opt}& Optical V surface brightness analysis & 42275 & (2),(3)\\ 
   GASPHOTB     & Sec. \ref{sec:opt}& Optical B surface brightness analysis & 41463 & (2),(3)\\ 
   NIRK               & Sec. \ref{sec:nir} & Near infrared (K) photometry & 962344 & (4)\\
   NIRKJ               & Sec. \ref{sec:nir} & Near infrared (J) photometry & 628813 & (4)\\
   GASPHOTK       & Sec. \ref{sec:nir} & Near infrared (K) surface brightness analysis & 71687& (2), (3)\\   
   MORPHOT        & Sec. \ref{sec:morph} & Morphology & 39923& (5)\\
   REDSHIFT        & Sec. \ref{sec:redshift} & Redshift and membership & 6132&(6)\\
   SFHIST              & Sec. \ref{sec:sfhist} & Star formation histories & 5299&(7), (8)\\
   EQWIDTH         & Sec. \ref{sec:eqwidth} & Equivalent widths & 4381& (9)\\
   LOCDENS        & Sec. \ref{sec:locdens} & Local densities & 66164 &--\\
 \hline                                   
 \end{tabular}
\tablebib{(1) \citet{Varela2009}; (2) \citet{Pignatelli2006}; (3) Bindoni et al., in preparation; (4) \citet{Valentinuzzi2009}; (5) \citet{Fasano2012}; (6) \citet{Cava2009}; (7) \citet{Fritz2007}; (8) \citet{Fritz2010}; (9) \citet{Fritz2014}
}
\end{table*}

\begin{figure*}
   \centering
   \includegraphics[width=0.85\textwidth]{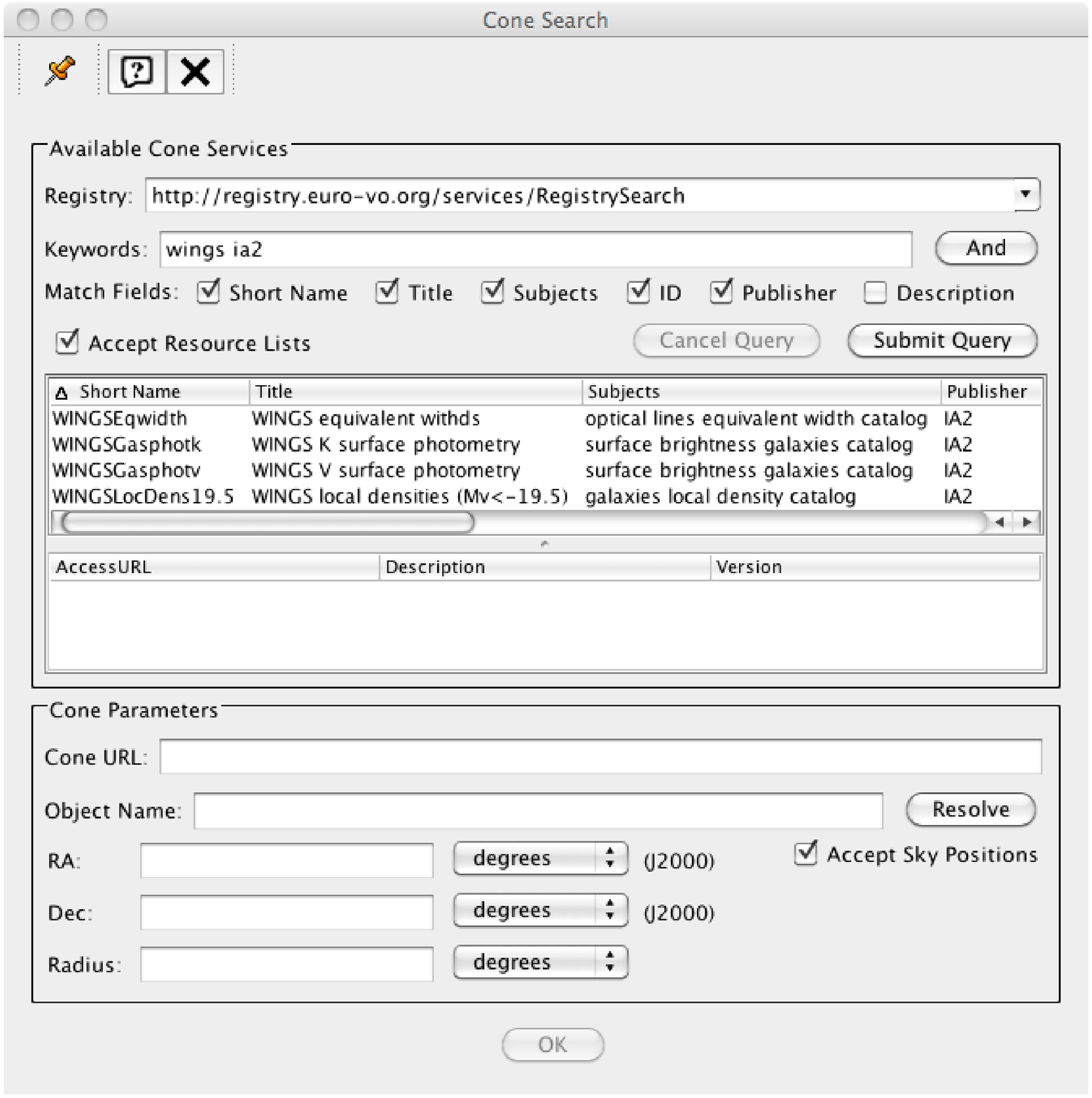}
      \caption{A TOPCAT screenshot of the cone--search query in the VO using the {\it wings ia2} keywords.}
         \label{Fig:voquery}
   \end{figure*}

Each one of the WINGS tables described in the present paper is available to the community as a cone--search in the VO framework \citep{Williams2011}. The registry to be looked at is the VO-registry ({\it http://registry.euro-vo.org/services/RegistrySearch}), and the available dataset can be found by using the keyword WINGS. Those whose field {\it Publisher} is IA2 are described in this paper and updated in real time when needed.
As a future development we will make available the entire relational database as a TAP service \citep{Dowler2011}, to allow more complex queries.
Fig. \ref{Fig:voquery} shows how to access WINGS data using the TOPCAT interface and the described keywords.


\section{Optical catalogs: photometry and structural parameters [WINGS-OPT]}\label{sec:opt}
\begin{figure*}
   \centering
   \includegraphics[width=\textwidth]{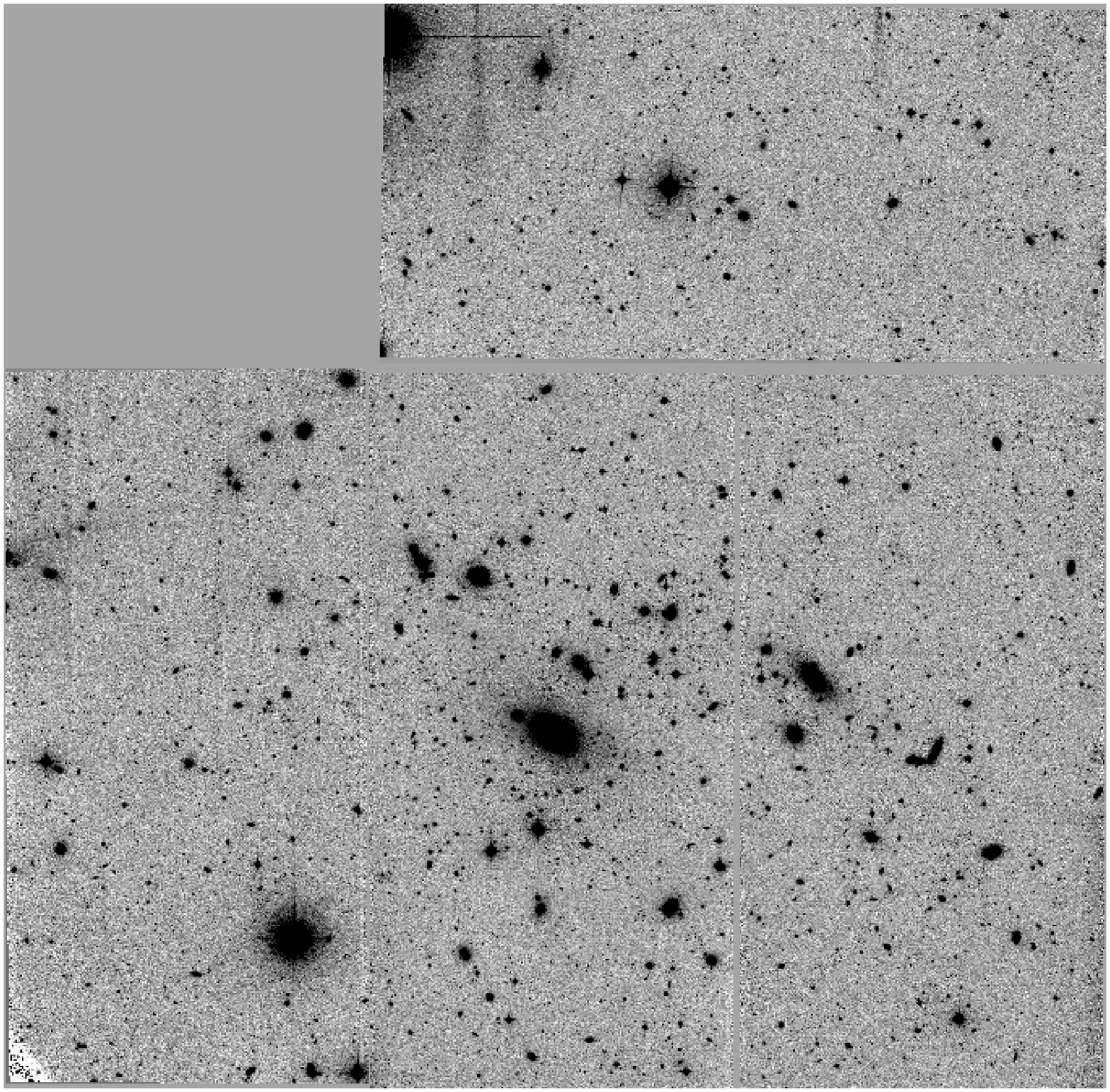}
      \caption{INT image of the cluster A85 in the V band as it is available for download as SIAP from the VO.}
         \label{Fig:a85_V}
   \end{figure*}

The core of the WINGS project is the optical photometry based on wide--field images of galaxy clusters at low redshift.
The wide--field images were taken using the wide field cameras on either the 2.5 m Isaac Newton Telescope (WFC@INT) or the MPG/ESO-2.2 m telescope (WFI@ESO). The typical FWHM for the whole set of 6 observing runs (3 for each telescope) was $\sim 1.1$ arcsec, while the magnitude limit was $M_V \sim -14$.
The mosaic images, whose dimensions are $\sim 35' \times \sim 35'$, are available as Simple Image Access Protocol (SIAP) \citep{Tody2011} through the VO tools (see Fig.\ref{Fig:a85_V} for an example). 
The IVOA identifier is {\it ivo://ia2.inaf.it/hosted/wings//opt}.
The pixel scale is 0.333 arcsec/pix for the INT images, and 0.238 arcsec/pix for the WFI images.
It is these images from which the WINGS photometry measurements were made.  These are described in \citet{Varela2009}, and we also refer the reader to this paper for an assessment of the image quality. The mosaic construction is also described in \citet{Fasano2006}.
The header of each downloadable image contains also the photometric zeropoint.

Photometry was performed on the V-band images using Sextractor \citep{Sex}. The V-band image was used as a reference for the subsequent B-band photometry (run in single and in double image mode). At the end of the process only objects detected in both bands were kept.
We detected a total  of 759,024 objects in 76 clusters, of which 394,280 are classified as galaxies in the V-band, 183,792 as stars and the remaining as unknown. 
The WINGS classification (see Tab. \ref{table:opt}) was mainly based on the Sextractor stellarity index and, as a starting point, we chose rather conservative limits: objects with stellarity index smaller than 0.2 were initially flagged as galaxies, those with stellarity larger than 0.8 as stars, and objects in between as unknown. 
The final classification, however, took into account other diagnostics, and we refer the reader to the original paper \citep{Varela2009} for a comprehensive description.
Unknown objects start to contaminate the galaxy sample starting from $V=21$, while at brighter magnitudes the fraction of misclassifications is negligible. At $V=22.5$ the fraction of unknown objects rises to $\sim 20\%$, becoming similar to the fraction of stars. A statistical study, however, demonstrates that the star/galaxy classification holds up well to $V\sim24$ \citep[see Fig. 7 and 8 in ][]{Varela2009}.
The published table contains the WINGS unique ID, as well as its sky coordinates and the cluster name, the classification (1 for galaxies, 2 for stars and 0 for unknown objects), the Sextractor internal binary flag that is useful to assess at a glance the photometric quality of the detected object (i.e. if it has close neighbors, if it is blended, and so on), the isophotal area above the analysis threshold, the ellipticity, position angle both from Sextractor and the usual one (N/E), and the peak surface brightness above the background.
Magnitudes given in the database were estimated at various physical apertures, namely 2.15 kpc, 5.38 kpc, 10.77 kpc at the cluster redshift, calculated using a standard cosmology of ($\Omega_M$=0.3, $\Omega_{\Lambda}$= 0.7, and H$_0$=70 km s$^{-1}$Mpc$^{-1}$). We also give aperture magnitudes for three fixed angular apertures of 1.6, 2.0 and 2.16 arcsec, corresponding to the fibre diameters of our spectroscopic observations.
All magnitudes were calibrated taking into account color equation and corrected for atmospheric absorption.
Errors in the given magnitudes were calculated using the average relation shown in Fig. 3 of \citet{Varela2009}.
We also list three aperture colors, derived from the set of three aperture magnitudes.
Finally we give the distances of the given object from the optical center of its host cluster, the distance from the X--ray center and the distance from the BCG, all in arcsec.
The maximum error is 0.08 mag in the V-band for sources brighter than $V=20.5$ and 0.1 mag in the B-band down to $B=21$. 
As an indicator of the color gradient, we also quote the $(B-V)$ color at 5 kpc (see sect.\ref{sec:sfhist} for further details).

\begin{table}
\caption{OPT table content}            
\scriptsize
\label{table:opt}      
\centering             
\begin{tabular}{l l l }
\hline\hline           
Content & Units & Description \\ 
\hline                        
   ID                       & --     & WINGS identifier (unique) \\ 
   RA                     & deg & Right ascension (J2000)\\ 
   DEC                  & deg & Declination (J2000) \\ 
   CLUSTER        & --     & Cluster name                        \\
   CLASS              & --     & WINGS classification (1, 2 or 0) \\
   SEX FLAG       & --     & Sextractor binary flag\\
   AREA                & arcsec$^2$ & Isophotal area above analysis threshold \\ 
   ELL                    & -- & Ellipticity ($1-b/a$) \\ 
   THETA              & deg & Sextractor position angle (CCW/x) \\ 
   PA                     & deg & Position angle (N/E) \\ 
   MU MAX         & mag/arcsec$^2$ & Peak surface brightness above background \\ 
   V AUTO           & mag & Kron--like elliptical aperture magnitude \\ 
   V 2KPC           & mag & Aperture mag,   R$=2.15  $kpc  \\ 
   V 5KPC           & mag & Aperture mag,   R$=5.38  $kpc  \\ 
   V 10KPC        & mag & Aperture mag,   R$=10.77  $kpc  \\ 
   V 160              & mag & Aperture mag,   D$=1.60  $arcsec  \\ 
   V 200              & mag & Aperture mag,   D$=2.00  $arcsec  \\ 
   V 216              & mag & Aperture mag,   D$=2.16  $arcsec  \\ 
   V AUTO ERR & mag & Error on V mag from \citet{Varela2009}\\
   B AUTO           & mag & Kron--like elliptical aperture magnitude \\ 
   B 2KPC           & mag & Aperture mag,   R$=2.15  $kpc  \\ 
   B 5KPC           & mag & Aperture mag,   R$=5.38  $kpc  \\ 
   B 10KPC        & mag & Aperture mag,   R$=10.77  $kpc  \\ 
   B 160              & mag & Aperture mag,   D$=1.60  $arcsec  \\ 
   B 200              & mag & Aperture mag,   D$=2.00  $arcsec  \\ 
   B 216              & mag & Aperture mag,   D$=2.16  $arcsec  \\ 
   B AUTO ERR & mag & Error on B mag from \citet{Varela2009}\\
   BV 2KPC       & mag & Aperture color $B-V$,   R$=2.15  $kpc\\
   BV 5KPC       & mag & Aperture color $B-V$,   R$=5.38  $kpc\\
   BV 10KPC       & mag & Aperture color $B-V$,   R$=10.77  $kpc\\
   D$_{OPT}$                 & arcsec  & Distance from the cluster optical center \\
   D$_{XCEN}$              & arcsec & Distance from the cluster X--ray center \\
   D$_{BCG}$                 & arcsec & Distance from the cluster BCG \\
\hline                                  
\end{tabular}
\end{table}

On the same mosaic images used to perform the photometry we also ran the GASPHOT tool \citep{Pignatelli2006} to analyze the surface photometry of WINGS galaxies and to derive their structural parameters. We derived such measurements for 42,278 galaxies in the V-band and for 41,463 galaxies in the B-band.
The complete procedure, as well as an assessment of the performance of GASPHOT will be described in Bindoni et al., in preparation.
In brief, the tool is designed to run in a completely automatic mode and basically fits a single Sersic law, simultaneously, to the light profiles along the major and the minor axis of each galaxy, appropriately convolved with a position varying point spread function. 
The software performs a fit of the growth profiles, so that the weight is higher where the uncertainties are smaller (i.e. in the galaxy external regions), and it is lower in the cores, where the galaxy photometry might be affected by peculiarities such as point-like sources, dust lanes, bars, pseudo-bulges and so on.  
GASPHOT measures the V magnitude, the effective radius, the mean surface brightness, the Sersic index and the axial ratio for each galaxy, as listed in Tab. \ref{table:gasphotv}. The errors on the derived quantities estimated for 90\% of the global sample are $<0.1$ mag for the V magnitude, $<0.2$ arcsec for the effective radius, $<0.8$ for the Sersic index, and $<0.015$ for the axial ratio.

Since each mosaic image has its own photometric quality, we flagged as bad measurements those which have errors in the derived parameters larger than the 98th percentile of the measurement errors in each cluster.
Furthermore, we also flagged as bad those fits which show Sersic indices at the limits of the space parameters (i.e. $n_V=0.5$ or $n_V=8$).
The quality flag is thus a binary number of 8 digits converted into decimal.
The first two digits are always 0, the remaining 6 are set to one when
the solution is extreme (i.e. Sersic index =0.5 or =8) [3rd digit], and when
the errors in the estimated parameters (magnitude [4th], effective radius [5th],
Sersic index [6th], background [7th] and axial ratio [8th]) exceed the 98th percentile of
the error distributions for the given cluster and filter.
Therefore, the flag is 0 for good fits, 32 for fits that find extreme solutions (i.e. Sersic index=0.5 or 8), 2 for fits with too large an error on the background estimation, and 16 for fits with too large error on the estimated magnitude.
The binary flag has been calculated in this way in order to recognize fits that are relatively bad, while we suggest to use the absolute errors to check for absolute deviant fits.

\begin{table}
\caption{GASPHOT table content: the same structure is present for the V, B and K measurements.}            
\scriptsize
\label{table:gasphotv}      
\centering                  
\begin{tabular}{l l l }     
\hline\hline                
Content & Units & Description \\
\hline                        
   ID                                & --     & WINGS identifier (unique) \\     
   RA                             & deg & Right ascension (J2000)\\ 
   DEC                          &  deg & Declination (J2000) \\ 
   CLUSTER                 & --     & Cluster name                        \\
   MAG GAS                        & mag &  magnitude \\
   MAG GAS ERR              & mag &  Error on the magnitude \\
   R$_{e}$                    &  arcsec & Effective radius (along semi-major axis)\\
   R$_{e}$ ERR           & arcsec & Error on the R$_e$\\
   $<\mu_{e}>$     & mag/arcsec$^2$ & Mean surface brightness \\
   N                       & -- & Sersic index \\
   N ERR             & -- & Error on the Sersic index \\
   AXRAT                      & -- & Axial ratio \\
   AXRAT ERR             & -- & Error on the axial ratio \\
   FLAG                         & -- & Quality flag \\
\hline                                   
\end{tabular}
\end{table}

\section{Near infrared catalogs: photometry and structural parameters [WINGS-NIR]}\label{sec:nir}
In order to better characterize WINGS cluster galaxies, in particular their stellar masses, without being biased by the latest star formation episode which dominates the galaxy light in the visible bands, WINGS is complemented by ancillary observations in the infrared bands taken with WFCAM@UKIRT.
A complete description of the data reduction and quality assessment  can be found in \citet{Valentinuzzi2009}.
We obtained observations in several runs ($\sim 14$) with a median seeing of $\sim 1.0$ arcsec, and the final photometric catalogues have a $90\%$ completeness of $19.4, 20.5$ in the K and J band, respectively.
Only a subsample of 28 clusters (17 clusters have been observed in both bands) have the needed photometric accuracy. These, however, cover the entire range of X--ray luminosities of our original sample of clusters. The redshift distribution is representative as well of the entire sample, while the sample is slightly  biased towards the low velocity dispersion tail of the cluster distribution.
The WFCAM data reduction and photometric calibration was performed by the CASU team \citep{Irwin2004,Hodgkin2009}
The area covered by observations corresponded to $\sim 0.79 deg^2$ for each cluster, thus making WINGS-NIR by far the largest survey of nearby galaxy clusters as far as the area coverage is concerned. In fact, in this redshift range, only individual clusters or small cluster samples have been studied in the literature up until now \citep[e.g.][]{Pahre1999,
Gavazzi1990,DePropris2003}. The WINGS survey of near-infrared data consists of 962,344 (628,813) sources in K (J) bands, of which 490,034 (263,116) are galaxies.
Star/Galaxy classification was based again on the Sextractor stellarity parameters, but we used a value of 0.35 (instead of 0.2, see sec.\ref{sec:opt}) to tag galaxies. This choice of parameters was done on the basis of the results of our artificial star tests.
This initial classification was then refined by using interactive cleaning of different populations in appropriate diagnostic plots. We refer the reader to the original paper by \citet{Valentinuzzi2009} for further details.
The structure of the NIRK and NIRJ tables of the WINGS database is analogous to that of the OPT table, and is shown in Tab. \ref{table:nirk}.
The VO table contains the unique WINGS ID, its sky coordinates, the cluster name, the WINGS classification, the WINGS binary flag (described below), the area, ellipticity and position angle (CCW/x), the peak surface brightness above the background, the Sextractor AUTO magnitude, as well as the aperture magnitudes inside 2.15, 5.38, 10.77 kpc and inside 1.60, 2.00, 2.16 arcsec. The errors on magnitudes are calculated following equation 7 in \citet{Valentinuzzi2009}. Finally we list for each object its distance in arcsec from the optical and X--ray centres and from the BCG.
The maximum error is 0.06 mag in both infrared bands for sources brighter than $16.5$.

\begin{table}
\caption{NIRK-NIRJ table content}             
\scriptsize
\label{table:nirk}     
\centering                         
\begin{tabular}{l l l }        
\hline\hline                 
Content & Units & Description \\ 
\hline                        
   ID                       & --     & WINGS identifier (unique) \\      
   RA                     & deg & Right ascension (J2000)\\ 
   DEC                  & deg & Declination (J2000) \\ 
   CLUSTER        & --     & Cluster name                        \\
   CLASS              & --     & WINGS classification (1, 2 or 0) \\
   WINGS FLAG & --     & WINGS binary flag\\
   AREA                & arcsec$^2$ & Isophotal area above analysis threshold \\ 
   ELL                    & -- & Ellipticity ($1-b/a$) \\ 
   THETA              & deg & Sextractor position angle (CCW/x) \\ 
   MU MAX         & mag/arcsec$^2$ & Peak surface brightness above background \\ 
   K,J AUTO           & mag & Kron--like elliptical aperture magnitude \\ 
   K,J 2KPC           & mag & Aperture mag,   R$=2.15  $kpc  \\ 
   K,J 5KPC           & mag & Aperture mag,   R$=5.38  $kpc  \\ 
   K,J 10KPC        & mag & Aperture mag,   R$=10.77  $kpc  \\ 
   K,J 160              & mag & Aperture mag,   D$=1.60  $arcsec  \\ 
   K,J 200              & mag & Aperture mag,   D$=2.00  $arcsec  \\ 
   K,J 216              & mag & Aperture mag,   D$=2.16  $arcsec  \\ 
   K,J AUTO ERR & mag & Error on K,J mag from \citet{Valentinuzzi2009}\\
   D$_{OPT}$                 & arcsec  & Distance from the cluster optical center \\
   D$_{XCEN}$              & arcsec & Distance from the cluster X--ray center \\
   D$_{BCG}$                 & arcsec & Distance from the cluster BCG \\
\hline                                   
\end{tabular}
\end{table}

The WINGS flag is calculated as follows
\begin{equation}
WINGS FLAG=a_1+2a_2+4a_3+8a_4
\end{equation}
where
\begin{itemize}
\item a$_1$=1 if classified as galaxy
\item a$_2$=1 if classified as star
\item a$_1$=a$_2$=0 if classified as unknown
\item a$_3$=1 if weakly affected by neighboring halo
\item a$_4$=1 if strongly affected by  neighboring halo 
\end{itemize}

\begin{figure*}
   \centering
   \includegraphics[width=0.65\textwidth,angle=90]{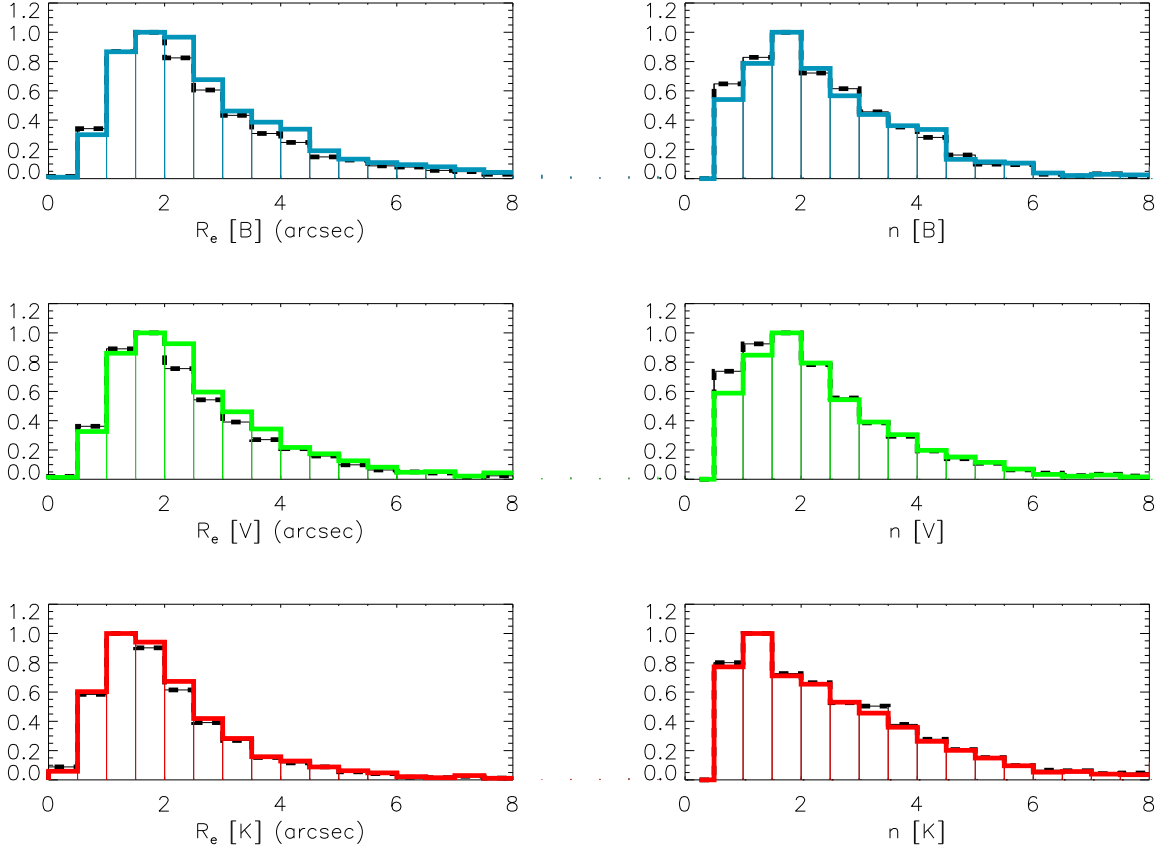}
      \caption{Distribution of Sersic index (right panels) and effective radius (left panels) retrieved by GASPHOT for the spectroscopic members of our WINGS clusters in the 3 bands covered by our photometry: from top to bottom data are from the B images, the V images and the K images. Dashed black lines refer to the entire WINGS sample, while the colored continuous lines are for the spectroscopic members observed in the three bands.}
         \label{fig:sersic_3bands}
   \end{figure*}

On the same K band images used to perform the photometric analysis we ran the GASPHOT code to determine the structural parameters of our sample of galaxies. This provided information for 71,687 galaxies. The corresponding table (called {\it gasphotk}) contains the same quantities described in Tab.\ref{table:gasphotv}.
For a subsample of 1254 galaxies, which are spectroscopic members of the WINGS clusters, we have the structural parameters in the 3 bands (B, V, K). This subsample is limited by the spectroscopic incompleteness (about 50\% of the observed targets turned out to be cluster members) and by the fact that not all clusters were actually observed in the K band.
Fig. \ref{fig:sersic_3bands} shows the distribution of the Sersic index (right panel) and of the effective radius circularized (left panel) for these galaxies as continuous lines (while the dashed lines in black correspond to the entire population, normalized to the same peak value).
As shown in the histograms of the effective radius, the light in K band is much more concentrated than in the visible bands, i.e. the mass (mainly traced by the infrared bands) is more concentrated than the light emitted in the visible bands, as expected. This obviously reflects also in the Sersic index distributions. The properties of member galaxies does not seem too much different from the overall distribution. The scientific interpretation of these distribution is going to be presented in a forthcoming paper.

\section{Morphology}\label{sec:morph}
The morphological classification of cluster galaxies is one of the most
important goals of the WINGS survey. Although visual classifications
are generally believed to be more reliable than any automatic classification
method, the advent of large mosaic CCDs has posed a big challenge to
galaxy classifiers, since it is impractical to visually
classify the tens of thousands galaxies imaged in large
surveys.

A remarkable effort to acquire visual classifications of huge galaxy
samples has been done by the Galaxy Zoo team \citep{Lintott2008,Lintott2011}, 
which recently made available the visual classifications for
about 900,000 galaxies from the Sloan Digital Sky Survey, as derived
from the contribution of more than 100,000 volunteers. However, 
the voluntary-based nature of these classifications makes them
necessarily coarse and only capable of distinguishing between elliptical
and spiral galaxies.

Many different approaches to automatic classification have been
proposed in the literature. Some of them have been widely used, as
large amounts of high quality data have become
available. Most automatic classifiers are based on morphological
proxies, like Concentration, Asymmetry, Clumpiness,
M20, the Gini coefficient, etc. \citep[see][among the others]{Abraham1996,Abraham2003,Conselice2003,Lotz2004,Lauger2005,Scarlata2007,VanderWel2008,Shamir2009,Cheng2011}.

The MORPHOT tool \citep{Fasano2012} combines a large set (21) of
diagnostics, easily computable from the digital cutouts of galaxies,
producing two different estimates of the morphological type based on:
(i) a semi-analytical Maximum Likelihood technique; (ii) a Neural
Network machine. The final estimator has been tested over a sample of
1,000 visually classified WINGS galaxies, proving to be almost as
effective as the 'eyeball' estimates themselves. In particular, at
variance with most existing tools for automatic morphological
classification of galaxies, MORPHOT has been shown to be capable of
distinguishing between ellipticals and S0 galaxies with unprecedented
accuracy.

The WINGS-MORPHOT catalog contains the morphological classifications of
39,923 galaxies, 2,963 of which have also been classified visually. The
numerical code adopted for morphology closely follows the revised
Hubble type classification, apart from cD galaxies being given a tag of 
$-6$ by MORPHOT, instead of $-4$. The final morphological type
given in the catalog is that from the visual classification if available, 
while in all other cases it is the mean of the two estimates provided by the
Maximum-likelihood (ML) and Neural Network (NN) techniques \citep[see][for further details]{Fasano2012}.

For each entry in the MORPHOT table, besides the unique WINGS ID, we
give (see Tab. 5) the sky coordinates, the cluster name, the
morphological types derived from the ML and NN techniques, together
with the relative confidence intervals, the mean morphological type
(again with the proper confidence interval), the visual type (when
available) and the final type that we propose to the astronomical
community for any scientific use. Details about the computation of the
confidence intervals are given in Appendix B of \citet{Fasano2012}.

\begin{table}
\caption{MORPHOT table content}             
\scriptsize
\label{table:morphot}      
\centering                 
\begin{tabular}{l l l }    
\hline\hline               
Content & Units & Description \\ 
\hline                        
   ID                       & --     & WINGS identifier (unique) \\      
   RA                             & deg & Right ascension (J2000)\\ 
   DEC                          &  deg & Declination (J2000) \\ 
   CLUSTER        & --     & Cluster name                        \\
   $T_{ML}$              & --     & Morphology from Max.likelihood technique \\
   $T_{ML}^{min} $ & --     & Lower limit of $T_{ML}$ \\
   $T_{ML}^{max} $ & -- &  Upper limit of $T_{ML}$\\ 
   $T_{NN} $             & -- & Morphology from Neural Network technique \\ 
   $T_{NN}^{min} $  & --&  Lower limit of $T_{NN}$ \\ 
   $T_{NN}^{max} $ & -- &    Upper limit of $T_{NN}$\\ 
   $T_{M}$                & --  & Morphology from both techniques\\ 
   $T_{M}^{min}$    & --  &   Lower limit of $T_{M}$\\ 
   $T_{M}^{max}$   & --  &   Upper limit of $T_{M}$ \\ 
   $T_{VIS}$            & --  &  Visual type (if any)\\ 
   TYPE               & --  &  Final MORPHOT type \\ 
\hline                                   
\end{tabular}
\end{table}

\section{Spectroscopic sample [WINGS-SPE]}\label{sec:spec}
Another primary goal of the WINGS survey was to produce a large dataset of galaxies in clusters with good quality spectra.
Due to the large field of view of the photometric sample, it has been possible to define candidates for the spectroscopic follow--up out to large distances from the cluster center (usually up to $\sim0.5\times R_{vir}$, but up to $R_{vir}$ in some cases).
The target selection took advantage of the WINGS photometric catalog (\citet{Varela2009}), already described in Sec. \ref{sec:opt}.
In order to maximize the probability of observing galaxies at the cluster redshift without biasing the cluster sample, targets were selected on the basis of their properties so that background galaxies (redder than the cluster red sequence) could be reasonably avoided. In particular the spectroscopic sample included only those galaxies with V $\leq 20$ (total magnitude), V$_{fiber} < 21.5$ and $(B-V)_{5 kpc} \leq 1.4$. This last cut was slightly varied from cluster to cluster in order to optimize the observational setup.
The number of targets with V $\leq 20$ is 30,126 in the global photometric sample, and 19,244 in the 48 clusters that were followed up spectroscopically, and it reduces to 28,861 (and 18,476, respectively) after having imposed the fiber magnitude cut.

Our total apparent magnitude limit (V $\sim 20$) is $1.5$ to $2.0$ mag deeper than the 2dFRS and Sloan surveys, and this is, in general, reflected in a higher mean number of member galaxies detected per cluster.
The spectroscopic observations were obtained over the course of 6 observing runs (22 nights) at the 4.2 m William Herschel Telescope (WHT) using the AF2/WYFFOS multifiber spectrograph and 3 observing runs (11 nights) at the 3.9 m Anglo Australian Telescope (AAT) using the 2dF multifiber spectrograph. 

In both cases the spectral range covers the optical range ($\sim 3800-7000$ \AA\ and $\sim 3600-8000$ \AA) where the most commonly used diagnostic lines are located (from the Ca H\&K in the blue to NaD in the red).
The dispersion was $\sim 6$ \AA\ and  $\sim 9$ \AA, while the nominal fiber aperture was 1.6" and 2", respectively.
Spectra for galaxies in only 48 of the original 76 clusters were obtained, due to bad weather conditions during the course of the observations, especially for the northern sample.

\subsection{Redshift and membership}\label{sec:redshift}
Redshifts were measured from the spectra using a semi-automatic method, which involves the automatic cross-correlation technique (as implemented in the {\it xcsao} IRAF task) and the emission lines identification. 

In order to determine which galaxies were cluster members, we used an iterative $\pm 3 \sigma$ clipping method \citep{Beers1990}, which allowed us to estimate the cluster velocity dispersion with an average number of galaxies that was up to 3 times larger than that used in previous studies of the same clusters. 
The spectroscopic completeness of the sample varies among the two datasets, being higher for the southern sample, where we have a spectroscopic completeness of 50\% at V=19.5 and could measure redshifts for 75\% of the galaxies. For the northern sample the situation is worse due to bad weather conditions during the observing runs \citep[see][for details]{Cava2009}.

The WINGS redshift table contains redshifts and memberships for 6,132 galaxies, 3,694 of which are tagged as cluster members.
Tab. \ref{table:redshift} shows the catalog entries.
The coordinates are those centered on the fiber, while errors on the redshift measurements are correlated with the amplitude of the correlation peak, as described in \citet{Cava2009} and references therein. The typical error is $\sim 25$ km/s.
The membership is set to 1 if the galaxy is considered to be a cluster member, otherwise it is set to 0.

\begin{table}
\caption{REDSHIFT table content}             
\scriptsize
\label{table:redshift}      
\centering                          
\begin{tabular}{l l l }        
\hline\hline                 
Content & Units & Description \\ 
\hline                        
   ID                       & --     & WINGS identifier (unique) \\     
   RA                    & deg & Right ascension (J2000)\\ 
   DEC                 & deg & Declination (J2000) \\ 
   cz		           & km/s & heliocentric velocity \\
   Err(cz)               & km/s & heliocentric velocity error\\
   z                         & -- & redshift \\
   membership    & -- & membership\\
\hline                                  
\end{tabular}
\end{table}

\subsection{Star formation histories}\label{sec:sfhist}
For the subsample of 5,299 galaxies with the highest signal--to--noise ratio ($\sim 15$) spectra, star formation histories were also derived via spectro-photometric modelling \citep{Fritz2010}.
In brief, a combination of Single Stellar Population models (SSPs) of different ages were used to derive the galaxy star formation history using a minimization technique. The SSPs spanned a range of ages between $10^6$ and $14.1 \times10^9$ years and were calculated for three different metallicities (i.e. Z/H = 0.004, 0.02 and 0.05). For each metallicity value, many realizations of the galaxy star formation history (SFH) were calculated, with the final result coming from the realization that minimised the difference between the observed and calculated quantities. 
In practice, the $\chi^2$ value was calculated as the weighted difference between the calculated and observed continuum fluxes and equivalent widths of the visible lines (both absorption and emission). The amount of dust was also a free parameter of the model which varied with the age of the SSP.

This approach does not take into account the chemical evolution of the galaxy, as it implicitly assumes that the overall population of the galaxy has just one, single metallicity. However, \citet{Fritz2007,Fritz2010} demonstrated that this assumption does not significantly bias the stellar mass determination. The adopted SSPs were those calculated using the isochrones of \citet{Bertelli1994} and a standard Salpeter IMF over the mass range 0.15--120 $M_{\odot}$.

The model was also used to compute galaxy stellar masses, based on the fiber aperture and total magnitudes (assuming no color gradient in the region between the fiber and the total extent of the galaxy), metallicity (intended as the metallicity of the best fit model), ages (both luminosity and mass weighted), average star formation rates in 4 main bins of age ($0-2\times10^7, 2\times10^7-6 \times 10^8, 6 \times 10^8-5.6 \times 10^9$ and $5.6 \times 10^9-14 \times 10^9$ years), as well as fiber and total (model) magnitudes in the whole range of observational filters (as detailed in the CDS version of the table).

The masses given in the table are of three types, as described in \citet{Fritz2007,Fritz2010} and references therein.
When transforming fiber masses to total masses the implicit assumption of no color gradient between the fiber aperture and the total galaxy magnitude is made.
However, we also have at our disposal the color gradient, since we measured B and V magnitudes at different distances from the galaxy centers. Therefore, we also list in the final table a term called {\it ccol}, computed using the fiber color and the color measured at 5 kpc from the galaxy center, which should be added to the total masses in order to take into account the color gradients.

In Tab. \ref{table:sfhist} we list for each of the 5,299 galaxies the metallicity of the best fit model, the age and total V--band extinction of the young stars calculated from the model, the star formation rates in four (broad but significant) bins of age, the stellar masses (inside the fiber and total) as (1) the mass of gas turned into stars, (2) the stellar mass, including mass locked up in remnants, and (3) the mass of stars still alive, the color term correction, the galaxy ages (both the luminosity and mass weighted).
The typical error on the masses is 0.2dex, while the maximum error on the age determination is $\sim 1$ Gyr (Fritz et al. 2011).

\begin{table}
\caption{SFHIST table content}             
\scriptsize
\label{table:sfhist}     
\centering              
\begin{tabular}{l l l } 
\hline\hline            
Content & Units & Description \\ 
\hline                        
   ID                       & --     & WINGS identifier (unique) \\      
   RA                    & deg & Right ascension (J2000)\\ 
   DEC                 & deg & Declination (J2000) \\ 
   metal                 & -- & best metallicity value\\
   A$_V$(young) & mag & V--band extinction of young stars \\ 
   A$_V$(tot)        & mag & Total V--band extinction \\
   sfr1                     & $M_{\odot}/yr$ & Star Formation Rate in the $0 - 2e7$ yr range\\
   sfr2                     & $M_{\odot}/yr$ & Star Formation Rate in the $2e7 - 6e8$ yr range\\
   sfr3                     & $M_{\odot}/yr$ & Star Formation Rate in the $6e8 - 5.6e9$ yr range\\
   sfr4                    & $M_{\odot}/yr$ & Star Formation Rate in the $5.6e9 - 14.1e9$ yr range\\
   m1$_{fib}$        & $M_{\odot}$  & Fiber Mass 1\\
   m2$_{fib}$        & $M_{\odot}$  & Fiber Mass 2\\
   m3$_{fib}$        & $M_{\odot}$  & Fiber Mass 3\\
   m1$_{tot}$        & $M_{\odot}$  & Total Mass 1\\
   m2$_{tot}$        & $M_{\odot}$  & Total Mass 2\\
   m3$_{tot}$        & $M_{\odot}$  & Total Mass 3\\
   ccol                    &  -- & color term \\
   LWAGE             &  yr  &  (Log10 of) Luminosity weighted age \\
   MWAGE            &  yr  &  (Log10 of) Mass weighted age \\
\hline                                   
\end{tabular}
\end{table}

\subsection{Equivalent widths}\label{sec:eqwidth}
A second set of measurements concerns the estimation of line equivalent widths (EW), that have been demonstrated to be a powerful tool to estimate stellar population properties (ages, star formation histories, metallicities and so on).

Line equivalent widths were measured in the  WINGS spectra using an automated method described in detail in \citet{Fritz2014}.
Among the original spectroscopic sample consisting of $\sim$ 6,000 galaxies, we discarded those having un-recoverable difficulties in the wavelength calibration shortwards of $\sim 4300$ \AA. Moreover, we also eliminated from the final sample clusters in which less than 20 objects turned out to be cluster members. At the end of the selection procedure, only 7 clusters from the northern sample possessed spectra that enter the final catalog of 4,381 objects, the remaining coming from the southern sample.

The measured spectral lines are listed in Tab. 1 of \citet{Fritz2014} and are listed in the database columns that we make publicly available (see Tab. \ref{table:eqwidth}). Following the WINGS ID with its sky coordinates, we list equivalent widths of OII, H$_{\theta}$, H$_{\zeta}$, H$_{\eta}$, Ca(K), Ca(H)+H$_\epsilon$, H$_{\delta}$, Gband, H$_{\gamma}$, H$_{\beta}$, OIII, Mg, Na, H$_{\alpha}$, D4000, D$_n$4000. We also give a classification (that is described below) and a completeness factor. 

\begin{table}
\caption{EQWIDTH table content}            
\scriptsize
\label{table:eqwidth}      
\centering                          
\begin{tabular}{l l l }       
\hline\hline                 
Content & Units & Description \\   
\hline                        
   ID                       & --     & WINGS identifier (unique) \\      
   RA                     & deg & Right ascension (J2000)\\ 
   DEC                  & deg & Declination (J2000) \\   
   OII                      & \AA & Equivalent width of OII [3727 \AA] \\
   Err(OII)              & \AA & Error on Equivalent width of OII [3727 \AA] \\
   H$_{\theta}$        & \AA & Equivalent width of H$_{\theta}$[3798 \AA] \\
   Err(H$_{\theta}$) & \AA & Error on Equivalent width of H$_{\theta}$ [3798 \AA] \\
   H$_{\zeta}$        & \AA & Equivalent width of H$_{\zeta}$[3889 \AA] \\
   Err(H$_{\zeta}$) & \AA & Error on Equivalent width of H$_{\zeta}$ [3889 \AA] \\
   H$_{\eta}$         & \AA & Equivalent width of H$_{\eta}$[3835 \AA] \\
   Err(H$_{\eta}$) & \AA & Error on Equivalent width of H$_{\eta}$ [3835 \AA] \\  
   Ca(K)                      & \AA & Equivalent width of Ca(K) [3934 \AA] \\
   Err(Ca(K))                     & \AA & Error on Equivalent width of Ca(K) [3934 \AA] \\
   Ca(H)+H$_\epsilon$                      & \AA & Equivalent width of Ca(H)+H$_\epsilon$ [3969 \AA] \\
   Err(Ca(H)+H$_\epsilon$)                     & \AA & Error on Equivalent width of Ca(H)+H$_\epsilon$ [3969 \AA] \\
   H$_{\delta}$         & \AA & Equivalent width of H$_{\delta}$[4101 \AA] \\
   Err(H$_{\delta}$) & \AA & Error on Equivalent width of H$_{\delta}$ [4101 \AA] \\  
   Gband                     & \AA & Equivalent width of Gband [4305 \AA] \\
   Err(Gband)                     & \AA & Error on Equivalent width of Gband [4305 \AA] \\
   H$_{\gamma}$         & \AA & Equivalent width of H$_{\gamma}$[4341 \AA] \\
   Err(H$_{\gamma}$) & \AA & Error on Equivalent width of H$_{\gamma}$ [4341 \AA] \\  
   H$_{\beta}$         & \AA & Equivalent width of H$_{\beta}$[4861 \AA] \\
   Err(H$_{\beta}$) & \AA & Error on Equivalent width of H$_{\beta}$ [4861 \AA] \\  
   OIII                      & \AA & Equivalent width of OIII [5007 \AA] \\
   Err(OIII)                     & \AA & Error on Equivalent width of OIII [5007 \AA] \\
   Mg                      & \AA & Equivalent width of Mg [5177 \AA] \\
   Err(Mg)                      & \AA & Error on Equivalent width of Mg [5177 \AA] \\
   Na                      & \AA & Equivalent width of Na(D) [5890+5895 \AA] \\
   Err(Na)                      & \AA & Error on Equivalent width of Na(D) [5890+5895 \AA] \\
   H$_{\alpha}$         & \AA & Equivalent width of H$_{\alpha}$[6563 \AA] \\
   Err(H$_{\alpha}$) & \AA & Error on Equivalent width of H$_{\alpha}$ [6563 \AA] \\  
   D4000               & \AA & Equivalent width of D4000 index (def. Bruzual 1983)\\
   D$_n$4000               & \AA & Equivalent width of D$_n$4000 index (def. Balogh 1999)\\
   Class                 & -- & Classification flag (see text)\\
   MagWeight      & -- & Magnitude weight (see eq.\ref{eqn:magcompl})\\
   RadWeight      & -- & Geometrical weight (see eq.\ref{eqn:radcompl})\\
\hline                                  
\end{tabular}
\end{table}

As for the completeness of the sample, both luminosity and geometrical completeness have been taken into account.
The total weight for each analyzed galaxy in the sample is defined as:

\begin{equation}
W(m,r)_i=\frac{1}{(C(m)_i \times C(r)_i)}
\end{equation}
where $C(m)_i$ and $C(r)_i$ are the magnitude and geometrical completeness in the bin to which the galaxy belongs.

The magnitude completeness accounts for the fact that not all the galaxies in each magnitude bin fulfill the selection criteria used for the spectroscopic sample.
The completeness as a function of magnitude is therefore defined as:
\begin{equation}\label{eqn:magcompl}
C(m) =\frac{N_z}{N_{ph}}(m)
\end{equation}
where N$_z$ is the number of galaxies with measured redshifts, and N$_{ph}$ is the number of galaxies in the parent photometric catalog, taking into account the cuts in color and magnitude, for each given magnitude bin m.

Moreover, we also computed the radial completeness for the WINGS sample, due to the fact that fiber collisions and superpositions are not allowed.
However, WINGS observations have been carried out with more than one configuration, with the result that the radial completeness function is close to being flat. In an analogous way to the magnitude completeness, the radial completeness is defined as:

\begin{equation}\label{eqn:radcompl}
C(r) =\frac{N_z}{N_{ph}}(r)
\end{equation}
where the bins are at varying radial distances from the center.

On the basis of these EW measurements (in particular [OII] and $H_\delta$) we also provide a spectral classification of our WINGS spectra, following the definitions given in \citet{Dressler1999}.
Broadly speaking, we classified as {\it e(a)}, {\it e(b)} and {\it e(c)} all emission line galaxies with supposedly high values of dust obscuration (value 1),
and stronger or weaker emission (values 2 and 3 in the database catalog).
Passive galaxies are instead labeled as {\it k}, {\it k+a}, {\it a+k}, the latter two classes showing the signature of recent but currently absent star formation (values 4, 5 and 6 of the catalog). 
When the [OII] emission line was not detectable, $H_\beta$ and [OIII] were used to distinguish emission line galaxies from passive galaxies.
A more detailed and careful description of the classification procedure, as well as an accurate description of our error evaluation, is given in \citet{Fritz2014}.

\section{Local densities}\label{sec:locdens}
The last piece of information added to the WINGS dataset is the local density of galaxies, that has been calculated in order to evaluate whether galaxy properties in clusters depend or not on the local environment. \citet{Vulcani2012} already used this quantity to study the dependence of the galaxy mass function on the environment.
Briefly, we recall here how the local density has been computed, and what is the meaning of the published columns of Tab. \ref{table:locdens} .
For each spectroscopically confirmed cluster member, the area of the circle containing its 10 nearest projected neighbors with photometry available and whose absolute magnitudes are brighter than $M_V=-19.5$, assuming them to cluster members, was calculated. 
Some of these galaxies will obviously be foreground or background field galaxies, and so these were subtracted statistically by using the counts given in \citet{Berta2006}.
A correction was also made to the local densities calculated for galaxies which lie at the edges of our WINGS images.
In these cases the area covered by the observations is smaller than the calculated area, their ratio being defined as the coverage factor.
The correction was applied by multiplying the number counts by this coverage factor and then doing the field galaxy subtraction. 
In both cases the area $A_{10}$ used was that obtained by interpolation between the two areas for which the corrected counts are immediately lower and higher than 10 (since the field number counts are not integer numbers in the canonical $A_{10}$ definition).
This computation has been made for all WINGS galaxies brighter than $M_V=-16$.
Note that for clusters lacking a velocity dispersion measurement (namely A311, A2665, A3164, and Z1261), the distances are given in pixels and not in terms of $R_{200}$.
In Tab.\ref{table:locdens} we list the data available in the VO, i.e. the local density of galaxies brighter than $M_v<=-19.5$  per Mpc$^2$ (LD), the percentage of area effectively covered by the observations (WLD), the distance from the X--ray center in units of $R_{200}$ or pixels (CCDX), and the percentage of circular area covered by the mosaic (WCCDX). These last two quantities have been calculated with respect to the BCG as well (CCDB and WCCDB, respectively).
\begin{table}
\caption{LocDens table content}            
\scriptsize
\label{table:locdens}     
\centering                        
\begin{tabular}{l l l }       
\hline\hline                 
Content & Units & Description \\    
\hline                        
   ID                       & --     & WINGS identifier (unique) \\      
   RA                     & deg & Right ascension (J2000)\\ 
   DEC                  & deg & Declination (J2000) \\   
   LD                      & N/Mpc$^2$ & (Decimal) Log of the nr. of galaxies per Mpc$^2$\\
   WLD                   & \%area & Coverage field fraction \\
   CCDX                & $R_{200}$/pix & Distance from the Xray cluster center\\
   WCCDX            & \%area & Coverage field fraction at CCDX \\
   CCDB                & $R_{200}$/pix & Distance from the BCG\\
   WCCDB            & \%area & Coverage field fraction at CCDX \\
 \hline                                   
\end{tabular}
\end{table}

\section{Summary}\label{sec:conclusions}
The WINGS survey has obtained data for 76 nearby clusters of galaxies. Here we describe and present the entire ensemble of WINGS catalogues that have been derived as part of this survey,  where the access key is the WINGS ID. 

In the near future we plan to produce new spectroscopic and photometric datasets for the outer regions of our cluster sample.
The data will be available through the VO tools and hosted at the Italian center for astronomical archives (IA2).

The released database contains optical (B, V) photometry for 759,024 sources (of which 394,280 are classified as galaxies) in 76 clusters at redshift $\sim 0.01-0.07$, with a maximum error of 0.08 mag in the V-band (to V$=20.5$) and 0.1 mag in the B band (to B$=21$). These data are supplemented by near infrared (J and K) photometry for 628,813 (in J) and 962,344 (in K) sources in 17 clusters (11 clusters have only the K-band photometry) with an error of  0.06 mag up to K,J$=16.5$.

We also provide measurements of the Sersic index, mean surface brightness, axial ratio, effective radius, and ellipticity for sub-samples of 42,278 galaxies in the V-band, 41,463 galaxies in the B-Band, and 71,687 galaxies in the NIR bands. Morphological classification are also given for 39,923 galaxies. Finally, we measured redshifts for 6,132 galaxies, of which 3,694 are cluster members. Errors on the redshift measurements are typically $25$ km/s.

Finally, two other catalogs are provided with information that complements the rest of the dataset: (i) a catalog containing star formation history information that includes masses, ages and star formation rates for 5,299 galaxies; (ii) a catalog giving equivalent width measurements for 4,381 galaxies. The errors on the derived masses are of the order of 10\% (median), and the maximum error on the ages is $\sim 1$ Gyr.
An estimate of the local galaxy density (for neighbours brighter than $M_V=-16$) has been calculated as well, taking into account the field coverage factors.

In order to download the data the user needs to access the VO-registry and use the keyword WINGS. Two types of data can be searched for: the wide--field images as a SIAP protocol, and scientific catalogs as a cone search. If accessing the VO with TOPCAT \citep{Taylor2005}, these two choices are listed under the {\it VO} tab. In order to access the data, the user must first select the 
appropriate registry ({\it http://registry.euro-vo.org/services/RegistrySearch}) and then insert the keywords WINGS and ia2 (just to avoid other catalogs with different publishers).
Once the resource has been selected, i.e after clicking on the line with the chosen source, other parameters become available, i.e. one can look for a particular object name, or particular position in the sky. In both cases the physical dimension of the search box must be specified. The typical screenshot is shown in Fig.\ref{Fig:voquery}. 
The results of these queries are a list of images that can be seen using related VO software (such as {\it Aladin}), for the SIAP, and the catalog table for the cone search.
Cross-matches between different tables/catalogs are possible, using the WINGS ID (i.e. the primary key of the database).
The entire relational database will be made available soon to the community using the TAP service of the VO \citep{Dowler2011}, thus allowing the database to be queried using the ADQL language.

Future releases of the database will include photometry in the U- and {\it u}-bands \citep{Omizzolo2013}, B- and V-band photometry for the outer cluster regions (based on VST observations), and analysis of spectra taken of galaxies in the outer regions with the AAT/AAOmega spectrograph.
\begin{acknowledgements}
BV is supported by World Premier International Research Center Initiative (WPI), MEXT, Japan
\end{acknowledgements}

\bibliographystyle{aa} 
\bibliography{wings_db.bib} 

\end{document}